# Overlapping Communities and the Prediction of Missing Links in Multiplex Networks

Amir Mahdi Abdolhosseini-Qomi [a,1], Naser Yazdani [a], and Masoud Asadpour [a]

[a] Department of Electrical and Computer Engineering, College of Engineering, University of Tehran, Tehran, Iran

*Abstract*—**Multiplex networks are a representation of real-world complex systems as a set of entities (i.e. nodes) connected via different types of connections (i.e. layers). The observed connections in these networks may not be complete and the link prediction task is about locating the missing links across layers. Here, the main challenge is about collecting relevant evidence from different layers to assist the link prediction task.**

**It is known that co-membership in communities increases the likelihood of connectivity between nodes. We discuss that co-membership in the communities of the similar layers augments the chance of connectivity. The layers are considered similar if they show significant inter-layer community overlap. Moreover, we found that although the presence of link is correlated in layers but the extent of this correlation is not the same across different communities. Our proposed, ML-BNMTF, as a link prediction method in multiplex networks, is devised based on these findings. ML-BNMTF outperforms baseline methods specifically when the global link overlap is low.**

*Index Terms*—**Multiplex Networks, Link Prediction, Inter-layer Community Overlap, Link Overlap, Bounded Non-negative Matrix Tri-Factorization, Constrained Non-convex Optimization, Coordinate Descent Method**

## I. Introduction

REAL world systems are made of elements with complex interconnections in between. Real-world systems like biochemical networks, human networks and air transportations are examples of biological, social, and technological systems, respectively. Scientists have studied these systems extensively under title of complex networks or network science [1][2]. The core concept of these researches is that the collective behavior of the whole system is not just a simple superposition of individual behavior of elements of the system [3]. These complex interactions lead to non-trivial behavior of the whole system. More specifically, neurons, human beings, and airports as the elements of the systems mentioned above are linked by inter-cellular connections, acquaintances, and flights respectively to shape the specific purposes of the systems [4][5][6].

Recently, a higher resolution image of these systems shows that the type of connection between elements of a system does not confine to one type but includes a variety of connection types [7][8][9]. Biological studies show that inter-cellular connections can be further divided into electrical and chemical connections [10]. Similarly, people are connected as they are members of a family, friends, or co-workers [5]. Also, a closer look into the air transportation system reveals that flights are not operated by a single airline, but dozens of airlines form the whole system [11]. So this new dimension of complexity may affect the behavior of complex networks, and it deserves to be studied with scrutiny.

The first step in this study is to find an appropriate mathematical representation for these systems. Multiplex networks are a suitable way of encoding this new dimension of complexity with a set of nodes which are connected in different layers via different types of link. This new representation, which is an extension to regular mono-layer (Simplex) networks, gives rise to the definition of a new set of structural properties, which are crucial to understanding the functionality of multiplex networks [12].

Real-world networks data is subject to incompleteness. One aspect of this incompleteness is the existence of missing links in the network [13]. The issue of multiplexity exacerbates the situation as knowing all types of connections among all entities requires more and more data, which is sometimes hard or impossible to be acquired. In biological studies, the discovery of different types of connections requires expensive lab experiments. In human networks, people are reluctant to expose some of their connections that are, for example, financially sensitive. Also, in the air transportation network, the rival airlines do not share the new flight openings. Therefore, the issue of finding missing links in the real-world multiplex network is a practically relevant problem.

---

[1] Corresponding author at Router and Social Network Analysis Labs, ECE department, University of Tehran
Tel.: +98-21-61114352
Email addresses: abdolhosseini@ut.ac.ir , yazdani@ut.ac.ir , asadpour@ut.ac.ir

The task of link prediction is about finding the missing links in the networks based on the structural similarity of unconnected node pairs [13]. The task of link prediction is related to the important task of community detection [14]. Communities are important as they describe the functional units of networks. From the link prediction point of view, the probability of linkage between two randomly chosen nodes in a sparse network sharply increases as we limit the set of nodes to those who belong to a community within the network. In other words, it is known that communities (and in a broader sense, hierarchies) are strong predictors of missing links in a broad spectrum of networks [14].

Finding missing links in a target layer of a multiplex network needs measures that quantify the structural similarity both from the view of the target layer and all other related layers. The main challenge is defining which layers are related and how is their relation to the target layer [15]. Here, we propose that layers which have similar community structure with a target layer should be considered as related, and the extent of relatedness differs from one community to another. Then the related communities of other layers can contribute to the link prediction task in the target layer in the same way that the communities of the target layer do.

The rest of this paper has been organized as follows: First, we review related works. Then, we report our observations, including the significance of inter-layer community overlap and variations of link overlap across communities in real-world multiplex networks. Later we describe our proposed method, ML-BNMTF, from a conceptual, mathematical and algorithmic point of view. Finally, we assess ML-BNMTF as a link prediction method for multiplex networks, and we conclude the work with a discussion about results and comparison with baseline methods.

II. RELATED WORKS

*A. Link prediction in Multi-relational and heterogeneous networks*

Around a decade of research in the link prediction has been focused on mono-layer (simplex) networks [16]. The extension of link prediction beyond simple networks has been studied under different names. In the machine learning community, this problem is known as Multi-relational Learning [17]. A multi-relational network is considered as a multiplex network in which the links are allowed to be directed. Most of the works in this area is based on heuristic loss functions with no connection to the structural properties of a multi-relational network. In the data mining community, the study of link prediction is done in heterogeneous networks. In heterogeneous networks, not only links but also nodes may have different types. Most of the methods in this domain are path-based, which is different from our approach.

*B. Link prediction in multiplex networks*

Recently, the problem of link prediction in multiplex networks has gained the attention of researchers. A systematic approach is extending the basic similarity measures to multiplex networks. However, when it comes to multiplex networks, it is hard to extend the notion of similarity [18].

Some researchers have approached the problem using feature engineering and applied machine learning. A study of a multiplex online social network, demonstrates the importance of multiplex links (link overlap) in significantly higher interaction of users based on available side information [19]. The authors consider the Jaccard similarity of an extended neighborhood of nodes in the multiplex network as a feature for training a classifier for the link prediction task. They have shown that using a multiplex feature enhances link prediction performance. Similar work on the same dataset benefits from node-based and meta-path-based features [20]. A specialized type of these meta-paths is tailored to be originated from and ending at communities. A binary classification has examined the effectiveness of the features for the link prediction task. Recently, other interlayer similarity features based on degree, betweenness, clustering coefficient, and similarity of neighbors have been used [21].

Furthermore, the issue of link prediction has been investigated in a scientific collaboration multiplex networks[22]. The authors have proposed a supervised rank aggregation paradigm to benefit from the node pairs ranking information, which is available in other layers of the network. Another study uses the rank aggregation method on a time-varying multiplex network [23]. The effect of other layers on the target layer of link prediction has been measured using a global link overlap. Recent work combines feature engineering and rank aggregation[24]. Two features based on hyperbolic distance are being used, and link overlap is considered for the relevance of layers.

The issue of layer relevance and its effect on link prediction is studied in [15]. The authors use global link overlap and the Pearson correlation coefficient of node features as measures of layer relevance, and later they use it to combine the basic similarity measures of each layer. The results support that the more layers are relevant, the better the performance of link prediction is attained.

*C. Community detection, overlapping communities and augmented linkage probability*

It is well-known that the structure of a network plays an important role in understanding the function of networks [1]. Modularity is one of the essential aspects of structural studies of networks that comprise concepts like communities [25]. Although still there is no consensus on definition of communities but regarding them as a set of nodes with high relative intra-community to inter-community links is a reasonable approach [26].

The main body of more than a decade of research on community detection is devoted to the partitioning of nodes into disjoint sets (non-overlapping communities). In [27], the authors revealed the significance of the overlap between communities in networks of nature and society. This issue led to a new line of research known as overlapping community detection [28][29][30]. Researchers have tackled this problem from different angles. Most important findings show that the density of links is higher in overlapping areas [31]. Also defining a global quality measure for overlapping communities is very challenging, and an escape way is using ground-truth information [32].

Real-world networks are mostly sparse, and the linkage probability is very low between two randomly selected nodes of these networks. This imbalance between links and non-links in the network makes link prediction a difficult task. Communities are parts of a network in which the probability of linkage increases sharply and hence why we can utilize them for enhancing link prediction performance. The fact that communities are good predictors of links [14], and they even do better when overlap with each other [31] are known concepts. Recently, an extension of Stochastic Block Models to multilayer networks has been done[33]. This work benefits from the model for link prediction task. Our work resembles the work of De Bacco et al. [34] as it studies community detection and link prediction tasks in multiplex networks at the same time. Differences between these two works can be outlined as follows:

- The membership vector in their model is the same for each node across all layers but ours is different for each layer.
- Their membership vector refers to overlapping multilayer communities but ours refers to overlapping layer-focused communities.
- For each layer, they infer an affinity matrix while we may infer additional affinity matrices for each layer w.r.t. the target layer of link prediction.

III. OBSERVATIONS

*A. A new dimension of overlaps*

Each layer of a multiplex network can be studied in separately from the community detection point of view. The communities in each layer may or may not overlap with each other. However, recent studies on social multiplex networks [35] show a new dimension of overlaps. New findings say that a non-random placement of communities exists in different layers of a multiplex network. To make a distinction, we call this new dimension of overlaps as Inter-layer community overlaps. To assess the idea of correlated presence of communities on different layers of multiplex network, we perform some experiments on datasets from real social, biological, and technological multiplex networks.

Indian Air-Train (AT) network is an example of interacting technological networks [36]. Sixty-nine geographically close airports and rail stations are the nodes of this multiplex network, and they are connected if a direct route exists between them, in each layer. Although air and train network are operated in different paradigms, but the spatial limitations impose some similar properties to these networks. On the other hand, the commuters benefit from both networks to reach to destination and hence why researchers are interested in studying the multiplex form of these networks.

C. Elegans (CE), Caenorhabditis elegans, is a small nematode, the first multicellular organism whose genome has been completely sequenced. The neural network of C.elegans, which is to date the only fully mapped brain of a living organism, is available for researchers [37][4]. The network consists of 281 neurons and around two thousand connections among them. An important aspect of this network, which is almost not considered in most of the analyses so far, is that the neurons can be connected either by a chemical link, a synapse, which can be further divided into Monadic and Polyadic links or by an ionic channel, the so-called gap junction, or electrical connections. These three types of connections have entirely different dynamics and function. Consequently, the neural network of the C. Elegans can be naturally represented as a multiplex network with 281 nodes and three layers, respectively, for electrical, chemical monadic, and chemical Polyadic junctions.

The Noordin Top Terrorist Network (NTN) Data were drawn primarily from "Terrorism in Indonesia: Noordin's Networks", a publication of the International Crisis Group [38], and include data on 79 individuals showing communication, financial, operational and trust relations between them as layers of a Multiplex network. Table 1 summarizes the main features of three Multiplex datasets under study in this work.

Now, for each layer of Multiplex networks under study we extract communities [39]. The outcome is sets of nodes grouped together as communities for each layer. The number of communities may vary from one layer to the other. To measure the level of overlap between communities of a pair of multiplex layers, we benefit from Normalized Mutual Information (NMI) [40]. NMI takes values in [0, 1] and its higher values indicates more similarity in community structure of the pair of layers [35]. Fig 1 shows the similarity of communities at different layers of NTN, CE and AT multiplex networks.

| Multiplex Network | N | M | $\langle n^{[\alpha]} \rangle$ |
|---|---|---|---|
| Air-Train (AT) | 69 | 2 | 69 |
| C. Elegans (CE) | 281 | 3 | 267 |
| Noordin Terrorists Network (NTN) | 78 | 4 | 56.25 |

TABLE 1. Main features of Multiplex networks under study in this work. The parameters n, m and $\langle n^{[\alpha]} \rangle$ represent the number of nodes, number of layers, and the average number of active nodes (non-isolated nodes) in multiplex networks, respectively.

The observations in Fig 1 indicate that inter-layer community overlap exists in real multiplex networks, but the extent of overlap may vary across different domains. Specifically, social multiplex networks show a higher level of overlaps compared to other domains like biological and technological domains.

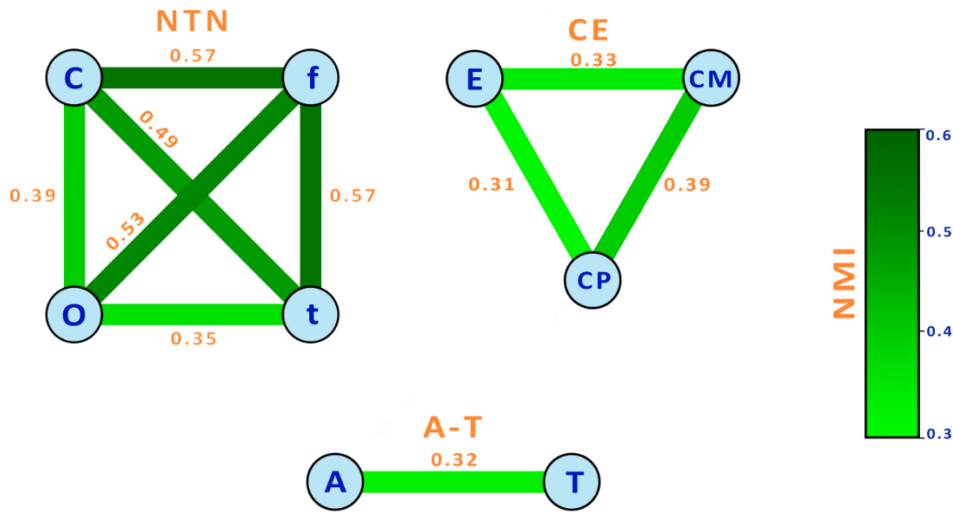

Fig 1- Similarity of community structure at different layers of networks under study. Every node represents a layer of a multiplex (NTN- C: Communication, O: Operation, F: Financial, T: Trust), (CE- E: Electrical, CM: Chemical-Monadic, CP: Chemical-Polyadic), (AT- A: Air, T: Train). The links are weighted according to Normalized Mutual Information between community structures of respective layers. The visualization idea is borrowed from [35].

In addition to the measurement of community overlaps, we have to determine the significance of this observation. Consider a multiplex network of two layers and n nodes. This multiplex is made by n identical one-to-one relations between nodes of two layers. So we can couple these two layers in n! different ways. All these realizations of a two-layer multiplex network preserve the internal structure of layers (including community structure) but alter the inter-layer measures like NMI. The observed NMI is significant if it is unlikely to see that NMI or extremer, subject to the random coupling of two layers. Using a hypothesis testing framework, we consider the Null Hypothesis (H0) as "the layers of multiplex are randomly coupled". For example, to test H0 on the AT network, we calculate the empirical CDF of NMI for ten thousands of random couplings of AT network layers (as shown in Fig 2). Then we have p-value = P(observed or higher NMI | H0) = 0. The result strongly rejects the Null Hypothesis. It means the observed NMI is not the result of a random coupling of layers, and it deserves to be considered for more analysis.

We conducted the same experiment for all layer pairs of datasets and results support the significance of the observed NMIs. The only exception is the Financial layer in the NTN dataset that we are not able to reject the Null Hypothesis concerning any other three layers.

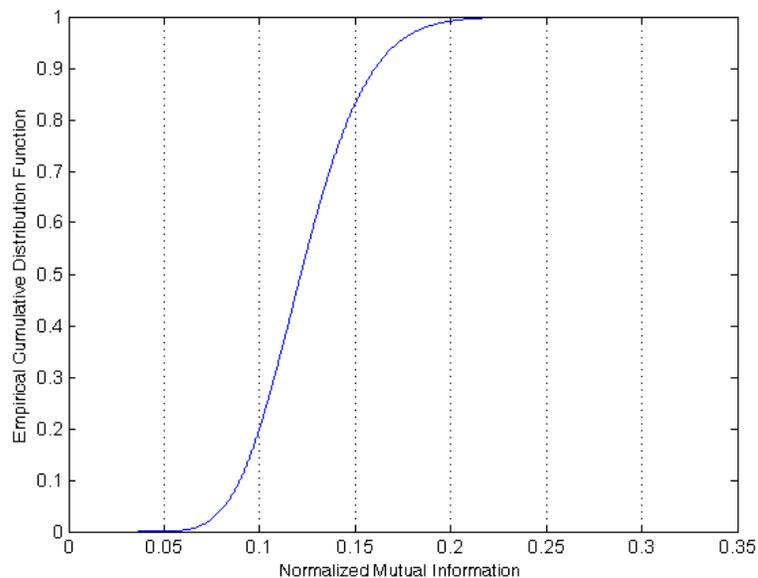

Fig 2- Empirical CDF of Normalized Mutual Information of ten thousand random couplings of AT multiplex network layers.

### B. Variations of edge overlap across different communities

Studies indicate that edge overlap is a significant observation in real multiplex networks[36][41]. Edge overlap is the conditional probability of finding a link at layer $\alpha'$ given the presence of a link between the same nodes at layer $\alpha$ [12]. In other words, this measure indicates how much seeing a link between two nodes in a layer gives us evidence about the possibility of seeing a link in another one. While this measure can be considered as a global edge overlap, we observe that the value of edge overlap varies across different localities defined by layers' communities. Fig 3 demonstrates the values of global link overlap and link overlap across different communities for sample duplexes from the real multiplex networks.

The variations of link overlap across different communities indicate that although the presence of links is correlated between layers of real multiplex networks but the extent of this correlation is subject to fluctuations. ML-BNMTF model can leverage these fluctuations to distinguish between weak vs. strong evidence of linkage in the target layer. Although, ML-BNMTF is not restricted to work only based on the direct overlap of links between layers, these observations give insight about the root of improvement in link prediction performance using this model.

Based on these observations, we continue with proposing a new factorization model called Multi-Layer Bounded Nonnegative Matrix Tri-Factorization (ML-BNMTF), which benefits from the correlated placement of communities across layers of a multiplex network.

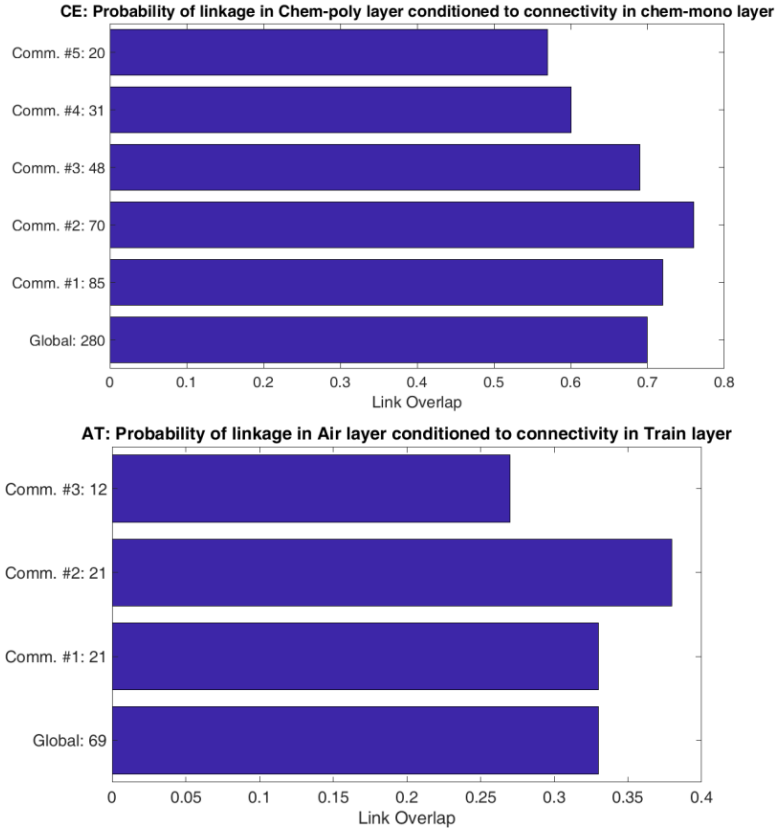

Fig 3- Global link overlap and variation of link overlap across communities in real multiplex networks. The bottom row in each plot shows the value of the global link overlap. In C. Elegans duplex (made of chem-poly and chem-mono layers), the link overlap for the top 5 communities of chem-mono layer and the number of nodes in each community is shown and the same is done for the top three communities of Train layer in Air/Train duplex network.

IV. ML-BNMTF

A. *Core concepts of ML-BNMTF*

Community overlap is a significant observation in networks of nature and society [27]. For example, in social networks, people are engaged in multiple friendship circles [5], and proteins in biological networks belong to several protein complexes [42]. Researchers have investigated the area of overlapping communities with different approaches [30].

More recently, methods based on different variations of matrix factorization have been proposed for the detection of overlapping communities. For example Nonnegative Matrix Factorization (NMF) [43] has been utilized for overlapping community detection [44][45], which has led to promising results. In addition, high scalability, good performance, and clear physical interpretation of the results make these methods more appealing [31].

In [46], authors have proposed a Bounded Nonnegative Matrix Tri-Factorization (BNMTF), which not only outperforms other rival NMF-based methods but also gives a clear physical interpretation of learned parameters. As shown in Fig 4a, the Tri-factorization method breaks the adjacency matrix $A$ of a simplex network of $n$ entities into three components, namely $U$, $B$, and $U^T$ which the latter is the transpose of the former. In other words, the adjacency matrix of a network can be estimated by the matrix product of these three components, and we have $A \approx UBU^T$. Considering $k$ as given number of communities in the network, the $U$ part is a $n \times k$ matrix with entries bounded between 0 and 1 and can be interpreted as membership probability of nodes in communities. Because the model allows membership in multiple communities, it is able to discover overlapping communities. According to the definition proposed in [47], a graph cover is a set of clusters such that each node is assigned to at least one cluster.

The graph cover is an extension of the traditional concept of graph partition (in which each node belongs to a single cluster), to account for possible overlapping communities. Therefore, the binary version of the matrix $U$ in which the nonzero elements are set to 1 will be a cover for the network under study. Then we have a B part which is non-negative and interpreted as interaction among all communities. This method of factorization is attractive as it covers the networks with weight and direction using augmented and asymmetric B, respectively.

Community detection meets new challenges in multiplex networks [48] as definitions are diverse, and the lack of ground-truth makes it difficult to evaluate proposed algorithms. The communities extracted by conventional community detection algorithms on each layer of a multiplex can be regarded as monoplex-communities and are referred to as auxiliary-partition. Most of the multiplex-community detection algorithms are based on post-processing of auxiliary-partitions related to each layer to extract meaningful multiplex partitions. In multiplex partitions, multiple types of links connect nodes.

Here, we propose Multi-Layer Bounded Non-negative Matrix Tri-Factorization (ML-BNMTF), which is different from earlier works as it assigns multiple auxiliary-partitions (or auxiliary-covers as the partitions are allowed to overlap) to each layer of a multiplex network. These extra auxiliary-covers come from other layers with relevant community structures. Considering two layers of a multiplex network with significant inter-layer community overlap, the communities of one layer are relevant to the other layer as far as they rebuild the structure of that layer. The extent of relevance should be learned as we elaborate in the following.

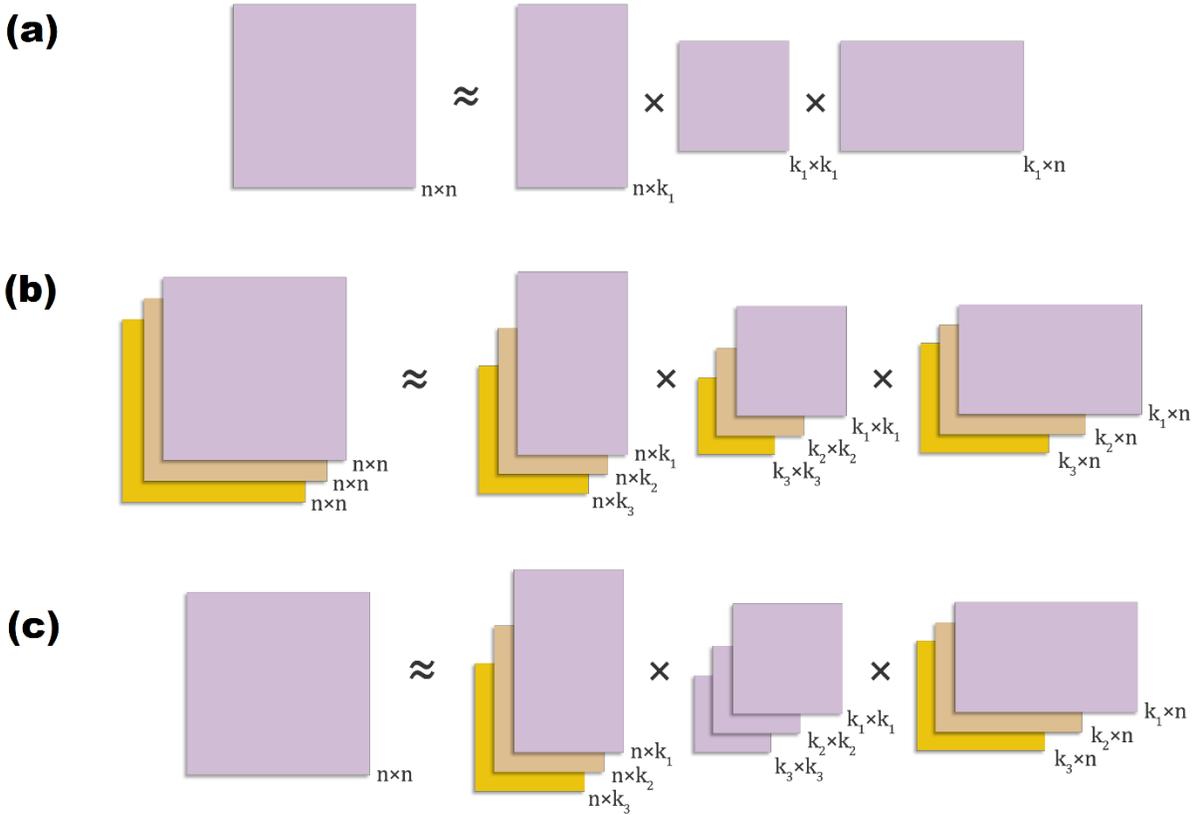

Fig 4- Illustration of factorization methods. (a) The Tri-factorization of an adjacency matrix, (b) Independent Tri-factorization of each layer of a multiplex, and (c) the proposed ML-BNMTF (Multi-layer Bounded Non-negative Matrix Tri-factorization) method for a target layer of link prediction. Note that all products are matrix-matrix products. The Adjacency matrix is of size $n \times n$. The given number of communities in the top, middle and bottom layers are $k_1, k_2$ and $k_3$ respectively.

Consider a multiplex network $G$ of $n$ nodes and $m$ layers as a set of adjacency matrices $\{A^1, A^2, ..., A^m\}$. It is obvious that for each layer of this multiplex, we can Tri-factorize the respective adjacency matrix (as shown in Fig 4b). The result will be in the form of $A^\ell \approx U^\ell B^\ell U^{\ell T}$, $\forall \ell \in \{1, ..., m\}$. Each layer may have a different number of communities. The physical interpretation of this setting is that every layer of a multiplex has its own community structure $(B^\ell)$ and nodes show independent affinity $(U^\ell)$ to these communities at each layer of the multiplex.

The significant inter-layer overlap of communities in real-world multiplex networks is the sign of the similarity of communities between layers. This fact, allows us to not only rebuild the structure of a target layer $h$ with its own communities $U^h$ but also with the communities of layers with similar communities. It should be considered that the communities of layers with similar communities will contribute to the rebuilding of the target layer but not in the same way as they do for their own layer. Therefore, as it has shown in Fig 4c, for each layer $\ell$ a matrix $B^{h,\ell}$ will be learned that determines the importance and interplay of communities of this layer when they contribute to the rebuilding of the target layer.

ML-BNMTF covers inter-layer, and intra-layer communities overlap simultaneously and can explore new dimensions of complexity in multiplex networks. As illustrated in Fig 5, the red and green communities are not the same in the top and the bottom layers, but they have high inter-layer overlaps. In other words, the inter-layer community overlaps determine some parts in the top bottom layer of the network in which the presence of links is correlated with each other. The extent of these correlations varies across different communities, as we discussed in Section III.B. We can infer that the shared nodes are playing a role in the emergence of these correlations. In addition, the nodes in one layer are not restricted to participate in only one community, as this is the property of BNMTF as an overlapping community detection method.

Considering the top layer as the target layer and the bottom layer as an auxiliary layer, ML-BNMTF looks for the role of red and green communities in the auxiliary layer in describing the seen links in the target layer. For those links that are less described by the communities of the target layer, conforming to communities of auxiliary layers means a new source of information is there that helps us to understand the structure of the target layer more. This approach gives us the ability to figure out the unseen links (missing links).

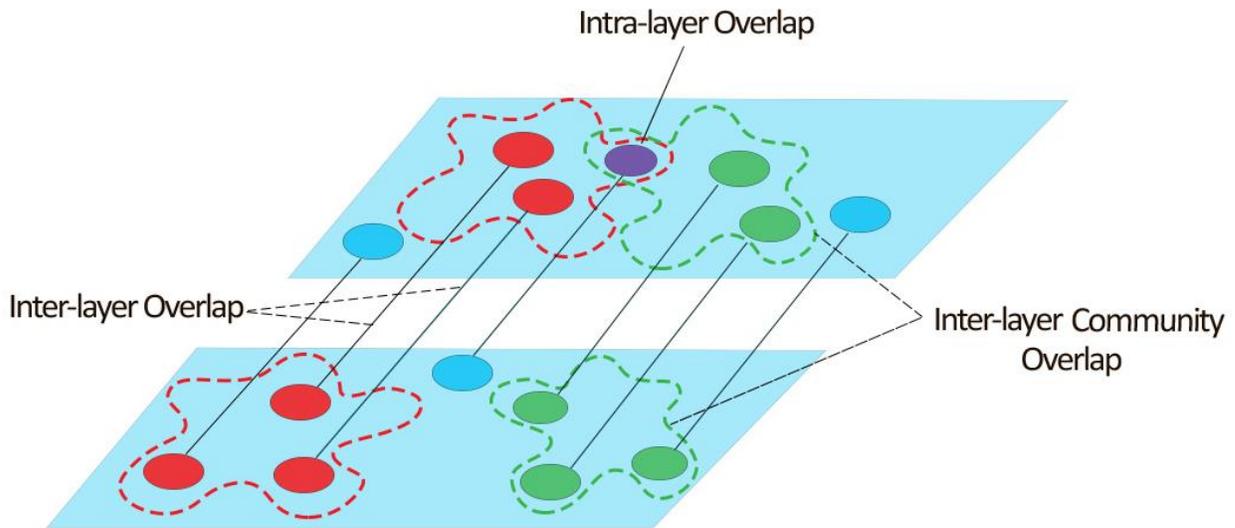

Fig 5- Illustration of ML-BNMTF capability of capturing intra-layer and inter-layer overlaps.

## B. Math of ML-BNMTF

For ease of comparison, we follow the same notation as [46] as much as we can. As mentioned earlier, we have a multiplex $G$ of $n$ nodes and $m$ layers as a set of adjacency matrices $\{A^1, A^2, ..., A^m\}$. So for $\ell^{th}$ layer we have $A^\ell \in \{0,1\}^{n \times n}$ whose (i, j)$^{th}$ entry $A^\ell_{ij}$ is equal to 1 if there exists a link between the i$^{th}$ and j$^{th}$ vertices on the $\ell^{th}$ layer of the multiplex network and zero otherwise. Also, we assume that for each layer $\ell$, the maximum number of communities, denoted by $k_\ell$, is given. For each layer, we use the matrix $U^\ell \in \mathbb{R}_+^{n \times k_\ell}$ to denote the community membership of the n vertices of that layer and $B^\ell \in \mathbb{R}_+^{k_\ell \times k_\ell}$ to denote the community interaction matrix. Each entry $u^\ell_{ij}$ in $U^\ell$ is the probability that i$^{th}$ vertex in layer $\ell$ belongs to the j$^{th}$ community in this layer. Therefore, we have the following constraint for $U^\ell$:

$$0 \leq u^\ell_{ij} \leq 1 \text{ or equivalently } 0 \leq U^\ell \leq 1 \quad (1)$$

The matrix $B^\ell$ shows the interaction between different communities in the layer $\ell$. The diagonal entries of $B^\ell$ show the strength of a specific community in the structure of the layer, while the off-diagonal entries indicate the strength of the relationship between different communities. Thus, for a target layer $h$ the product form $U^h B^h U^{h^T}$ rebuilds the corresponding adjacency matrix $A^h$ in terms of the communities and affinity of the vertices in this layer to them. BNMTF, as proposed in [46], solves the optimization problem

$$\min_{U^h, B^h} L_{sq}(A^h, U^h, B^h) \text{ s.t. } 0 \leq U^h \leq 1, B \geq 0 \quad (2)$$

to learn the appropriate $U^h$ and $B^h$, where

$$L_{sq}(A^h, U^h, B^h) = \|A^h - U^h B^h U^{h^T}\|_F^2 + \lambda \sum_{\ell=1}^{m}(1^T U^h 1) \quad (3)$$

Equation 3 benefits from squared loss function to measure the approximation error which is written in the form of Frobenius norm. Also, to avoid overfitting effect, which may happen when a node shows tendency to participate in many communities an additional term is considered. It is well known that $\ell^1$ norm as a regularization term has the ability to control the complexity of the model [49]. Adding the regularization term as in equation 3 (**1** is a matrix of all ones of the appropriate size), will lead to sparse Us and gives the ability to balance the tradeoff between the approximation error and the complexity of Us.

To solve the optimization problem, BNMTF uses the coordinate descent method. Coordinate descent methods update one variable at a time and are efficient at finding a reasonable solution. For $U^h$, Coordinate descent methods aim to conduct the following one variable updates:

$$(U^h, B^h) \leftarrow (U^h + tO^{pq}_{nk}, B^h) \quad (4)$$

where $O^{pq}_{nk}$ denotes a $n \times k$ matrix (here the appropriate size is $n \times k_h$) with all entries equal to 0 except the (p,q)th entry is equal to 1. Hence we have a one-variable optimization problem as

$$f^{U^h}_{pq}(t) = L_{sq}(A^h, U^h + tO^{pq}_{nk}, B^h) \quad (5)$$

$$\min_t f^{U^h}_{pq}(t)$$

$$s.t. \quad 0 \leq u^h_{pq} + t \leq 1$$

As any update of $U^h$ only affects our approximation of $A^h$ and has no impact on any other layer in the multiplex network, we can reuse the update rules mentioned in [46]. In update of matrix $B^h$, the symmetry of $A^h$ should be considered. The updates must keep $B^h$ symmetric, so there will be different objective functions for non-diagonal and

diagonal elements of $B^h$. The objective function for non-diagonal entries is given by

$$f_{pq}^{B^h}(t) = L_{sq}(A^h, U^h, B^h + tO_{kk}^{pq} + tO_{kk}^{qp}) \quad (6)$$

$$\min_t f_{pq}^{B^h}(t)$$

$$s.t. \quad b_{pq}^h + t \geq 0$$

In this way, we make sure that updates of $B^h$ do not change the symmetry of this matrix. Similarly, the update rule for diagonal entries of $B^h$ comes from solving the following one-variable optimization problem:

$$f_{pp}^{B^h}(t) = L_{sq}(A^h, U^h, B^h + tO_{kk}^{pp}) \quad (7)$$

$$\min_t f_{pp}^{B^h}(t)$$

$$s.t. \quad b_{pp}^h + t \geq 0$$

The corresponding update rules for diagonal and non-diagonal entries of $B^h$ matrix is mentioned in [46].

Now, let's consider a subset of layers $Z \subset \{1,...,m\} \setminus h$ with significant inter-layer community overlap with a layer $h$. The communities of each layer $\ell \in Z$ can contribute to rebuilding the structure of the target layer as

$$\tilde{A}^{h,\ell} = U^\ell B^{h,\ell} U^\ell \quad (8)$$

where $\tilde{A}^{h,\ell}$ is the rebuild of the target layer with communities of layer $\ell$ and $B^{h,\ell}$ determines the way that these communities contribute to the rebuilding process. The term $B^{h,\ell}$ is learned via solving the optimization problem

$$\min_{B^{h,\ell}} L_{sq}(A^h, U^\ell, B^{h,\ell}) \quad s.t. \quad B \geq 0 \quad (9)$$

The symmetric $A^h$ enforces $B^{h,\ell}$ to be symmetric as well. Thus for non-diagonal entries of $B^{h,\ell}$, the optimization problem is

$$f_{pq}^{B^{h,\ell}}(t) = L_{sq}(A^h, U^\ell, B^{h,\ell} + tO_{kk}^{pq} + tO_{kk}^{qp}) \quad (10)$$

$$\min_t f_{pq}^{B^{h,\ell}}(t)$$

$$s.t. \quad b_{pq}^{h,\ell} + t \geq 0$$

Here, the approximation error of the layer $h$ is:

$$L_{sq}(A^h, U^\ell, B^{h,\ell} + tO_{nk}^{pq} + tO_{nk}^{qp}) = \left\| A^h - U^\ell (B^{h,\ell} + tO_{nk}^{pq} + tO_{nk}^{qp}) U^{\ell^T} \right\|_F^2 + \lambda (1^T U^\ell 1)$$

Let's define $P_1^{h,\ell}$ and $P_2^{h,\ell}$ as follows:

$$P_1^{h,\ell} = U^\ell B^{h,\ell} U^{\ell^T} - A^h \quad (11)$$

$$P_2^\ell = U^\ell (O_{kk}^{pq} + O_{kk}^{qp}) U^{\ell^T} \quad (12)$$

Now we can rewrite the approximation error of layer $h$ as:

$$L_{sq}(A^h, U^\ell, B^{h,\ell} + tO_{nk}^{pq} + tO_{nk}^{qp}) = \left\| tP_2^\ell + P_1^{h,\ell} \right\|_F^2 + \lambda (1^T U^\ell 1)$$

Substituting $\left\| tP_2^\ell + P_1^{h,\ell} \right\|_F^2$ with the trace of $(tP_2^\ell + P_1^{h,\ell})^T (tP_2^\ell + P_1^{h,\ell})$, we have:

$$f_{pq}^{B^{h,\ell}}(t) = t^2 tr(P_2^{\ell^T} P_2^\ell) + 2t \times tr(P_1^{h,\ell^T} P_2^\ell) + tr(P_1^{h,\ell^T} P_1^{h,\ell}) + \lambda \sum_{\ell=1}^m (1^T U^\ell 1)$$

The optimal t subject to non-negativity of factors will be:

$$t = -\min(b_{pq}^{h,\ell}, \frac{tr(P_1^{h,\ell^T} P_2^\ell)}{tr(P_2^{\ell^T} P_2^\ell)}) \quad (13)$$

Similarly, the update rule for diagonal entries of $B^{h,\ell}$ comes from solving the following one-variable optimization problem:

$$f_{pp}^{B^{h,\ell}}(t) = L_{sq}(A^h, U^\ell, B^{h,\ell} + tO_{kk}^{pp}) \quad (14)$$

$$\min_t f_{pp}^{B^{h,\ell}}(t)$$

$$s.t. \quad b_{pp}^{h,\ell} + t \geq 0$$

In this setting, the approximation error of the layer $h$ is:

$$L_{sq}(A^h, U^\ell, B^{h,\ell} + tO_{kk}^{pp}) = \left\| A^h - U^\ell(B^{h,\ell} + tO_{kk}^{pp})U^{\ell^T} \right\|_F^2 + \lambda(1^T U^\ell 1)$$

Let's define $P_0^\ell$ as follows:

$$P_0^\ell = U^\ell O_{kk}^{pp} U^{\ell^T} \quad (15)$$

Now we can rewrite the approximation error of layer $h$ as:

$$L_{sq}(A^h, U^\ell, B^{h,\ell} + tO_{kk}^{pp}) = \left\| tP_0^\ell + P_1^{h,\ell} \right\|_F^2 + \lambda(1^T U^\ell 1)$$

Substituting $\left\| tP_0^\ell + P_1^{h,\ell} \right\|_F^2$ with the trace of $(tP_0^\ell + P_1^{h,\ell})^T (tP_0^\ell + P_1^{h,\ell})$, we have:

$$f_{pq}^{B^{h,\ell}}(t) = t^2 tr(P_0^{\ell^T} P_0^\ell) + 2t \times tr(P_1^{h,\ell^T} P_0^\ell) + tr(P_1^{h,\ell^T} P_1^{h,\ell}) + \lambda \sum_{\ell=1}^m (1^T U^\ell 1)$$

The optimal t subject to non-negativity of factors will be:

$$t = -\min(b_{pp}^{h,\ell}, \frac{tr(P_1^{h,\ell^T} P_0^\ell)}{tr(P_0^{\ell^T} P_0^\ell)}) \quad (16)$$

### C. Algorithm of ML-BNMTF

Having all update rules for variables involved, the algorithm of ML-BNMTF has been shown in Algorithm 1.

---

**Algorithm 1** ML-BNMTF

**Initialization:** choose the target layer $h \in \{1,...,m\}$ for link prediction

    Determine the set $Z$ of layers with significant inter-layer community overlap with target layer according to the procedure described in section III A

    random initialization of $U^\ell, \forall \ell \in Z \cup h$

    random initialization of $B^\ell, \forall \ell \in Z \cup h$

    random initialization of $B^{h,\ell}, \forall \ell \in Z$

**for** itr = 1 ,..., MAX_Iter **do**
  **for** $\forall \ell \in Z \cup h$ **do**
    Update $U^\ell$ according to equation 5 and update rules in [34]
    Update $B^\ell$ according to equation 6,7 and update rules in [34]
  **end for**
**end for**

```
for itr = 1 ,..., MAX_Iter do
    for ∀ℓ ∈ Z do
        Update $B^{h,\ell}$ off-diagonal entries according to equation 13
        Update $B^{h,\ell}$ diagonal entries according to equation 16
    end for
end for
return $U^{\ell}, \forall \ell \in Z \cup h$
       $B^{\ell}, \forall \ell \in Z \cup h$
       $B^{h,\ell}, \forall \ell \in Z$
```

## V. Assessment of ML-BNMTF

### A. Link prediction using ML-BNMTF

The link prediction algorithms are supposed to estimate the existence likelihood of all non-observed links based on the observed links of the network. For a multiplex network with a target layer with missing links, both the observed links of the target layer and the links of other layers should be exploited to achieve a better estimate of the existence likelihood of all non-observed links in the target layer.

Another way of representing a multiplex network is via sets. Consider a multiplex $G(V, E^1, ..., E^m)$ in which $V$ is a set of $n$ vertices and $E^{\ell} \subset V \times V, \forall \ell \in \{1,...,m\}$ is the set of links in a layer $\ell$. Denote by $U$ the universal set of all possible $\frac{|V| \times |V-1|}{2}$ links in the network, where $|V|$ is the number of elements in $V$. So, for the target layer $h$ the set $U - E^h$ will be the set of non-observed links of the network that contains the missing links which link prediction algorithms are supposed to locate them.

As the missing links are not known beforehand in real applications, to investigate the suitability of link prediction algorithms, some of the links in the target layer should be randomly removed to form the probe set $E_P^h$, and the remainder links are considered as the training set $E_T^h$. Obviously, $E_T^h \cup E_P^h = E^h$ and $E_T^h \cap E_P^h = \emptyset$. The link prediction algorithms for multiplex networks are allowed to use $E_T^h$ to locate $E_P^h$ among all possible choices in $U - E_T^h$, and if they do well, hopefully, they can do the same for missing links in $U - E^h$ for which there is no ground truth. For this purpose, the link prediction algorithms assign an existence likelihood score $S_{ij}^h$ to each non-observed link in the target layer $(i, j) \in U - E_T^h$.

For the target layer, ML-BNMTF gets a set of links in $E_T^h$ as input. So, $A^h$ will be the adjacency matrix that corresponds to links in $E_T^h$. Using $A^h$ and adjacency matrices of all layers in $Z$, ML-BNMTF returns the learned factors, as mentioned in Algorithm 1. Based on these factors, we propose the existence likelihood score

$$S_{ij}^h = U^h B^h U^{h^T} + \sum_{\forall \ell \in Z} U^{\ell} B^{h,\ell} U^{\ell^T} \quad (17)$$

for all non-observed links in the target layer.

The evaluation of the proposed and competing baseline methods is done based on Area Under the receiver operating characteristic Curve (AUC)[50]. This measure shows the probability that a randomly chosen missing link has a higher score than a randomly chosen nonexistent link. A good estimate of this measure can be achieved by sampling. A random sample from each of the missing links and nonexistent links is picked at each time. Considering $n$ independent samples out of which $n'$ times, the missing link has a higher score than the nonexistent link and $n''$ times they have the same score. Then the AUC can be calculated as:

$$AUC = \frac{n' + 0.5 \times n''}{n}.$$

Random assignment of scores leads to the AUC value of approximately 0.5. As the sorting according to existence likelihood score gets close to perfect sorting in which all missing links are scored higher than nonexistent links, the value of AUC approaches 1. In this way, the AUC measure evaluates the quality of the whole sorted list.

### B. Experiment on real-world datasets

Now, we report the outcome of applying ML-BNMTF to real multiplex datasets which are under study in this work. Empirical results show that the proposed coordinate descent method converges to local minima after a few iterations. We considered $MAX\_Iter = 6$. The increase of this number leads to slight improvement in the performance. In all experiments, the value of regularization factor $\lambda$ is constant and equals to 1. The number of communities in each layer is considered as given. Here, we derived this number by applying multiresolution community detection algorithm which gives a set of possible partition of the network [51]. Then, the partition which maximizes the modularity is selected and the respective number of communities is used [52]. Albeit, the issue of determining the number of communities in the BNMTF model is considered as future works by authors [46]. This can be done using the extended definition of modularity in overlapping communities as a future work [53].

The results of applying ML-BNMTF and three baseline methods are reported in Table 2. We conducted the experiment for the case of updating all entries of $B^{\ell}, B^{h,\ell}$ matrices and for the case that just updated diagonal entries of them and considering off-diagonals as zero. The results supported that using off-diagonal entries degrades the results, so we report the results of the second case. For all datasets which are under study, each possible pair of layers are considered as a real duplex network. The only exception is the Financial layer in NTN network which does not show a significant community overlap with other layers as we discussed in section III A. One of the layers of a duplex can be considered as target layer of link prediction and the other one as the auxiliary layer and vice versa.

| Dataset | Target layer | Auxiliary layer | BNMTF-T | BNMTF-T + Adjacency-A | BNMTF-T + BNMTF-A | ML-BNMTF |
|---|---|---|---|---|---|---|
| AT | Air | Train | 0.908 | 0.910 | 0.916 | **0.933** |
| AT | Train | Air | 0.838 | **0.851** | 0.839 | **0.851** |
| CE | Electrical | Chem-Mono | 0.559 | 0.642 | **0.791** | 0.786 |
| CE | Electrical | Chem-Poly | 0.565 | 0.672 | 0.807 | **0.808** |
| CE | Chem-Mono | Electrical | 0.834 | **0.852** | 0.841 | 0.839 |
| CE | Chem-Mono | Chem-Poly | 0.832 | **0.933** | 0.898 | 0.902 |
| CE | Chem-Poly | Electrical | 0.846 | **0.854** | 0.845 | 0.843 |
| CE | Chem-Poly | Chem-Mono | 0.847 | **0.902** | 0.886 | 0.885 |
| NTN | Communication | Operation | 0.875 | 0.904 | 0.885 | **0.908** |
| NTN | Communication | Trust | 0.871 | **0.916** | 0.896 | 0.913 |
| NTN | Operation | Communication | **0.987** | 0.976 | 0.984 | **0.987** |
| NTN | Operation | Trust | **0.986** | 0.945 | 0.954 | 0.983 |
| NTN | Trust | Communication | 0.913 | 0.945 | 0.946 | **0.949** |
| NTN | Trust | Operation | 0.911 | 0.883 | 0.882 | **0.937** |

TABLE 2. The evaluation of ML-BMNTF against baseline methods. The evaluation metric is AUC. Each row represents a sample duplex from real multiplex networks under study. Also, the target layer for link prediction and the auxiliary layer is specified. The probe set contains 10% of randomly selected links in the target layer. So, to avoid fluctuations due to random removal links, the average AUC over 100 implementations is reported. BNMTF-T column denotes applying BNMTF on the observed part of the target layer and scoring according to learned factors. The next column denotes the addition of Adjacency matrix to the score provided by BNMTF-T. The third baseline method applies BNMTF to both layers and uses the sum of the scores provided by each of them. The last column reports the results of the proposed ML-BNMTF method.

The results reported in Table 2 have several indications. The use of the information contained in the auxiliary layer can improve the performance of link prediction in the target layer. The only exception is when the Operation layer in the NTN network is considered as the target layer, and this can be attributed to the very high performance of applying BNMTF to the observed part of this network. In other cases, the best performance is divided between adding the adjacency matrix of the auxiliary layer to BNMTF-T (second method) and the proposed ML-BNMTF (fourth method). Our investigation shows that in the cases in which the global link overlap is high (like 0.7 in chem-poly/chem-mono in CE network as discussed in section III C), the second method wins but in the cases with low global link overlap (like 0.33 in Air/Train in AT network as discussed in section III C), ML-BNMTF outperforms. This is due to the fact that when global link overlap is above 0.5, the addition of the adjacency matrix of the auxiliary layer is more likely to promote missing links versus nonexistent links. In contrast, ML-BNMTF considers the nodes' affinity in the auxiliary layer to communities, $U^{\ell}$, and the communities' ability to rebuild the structure of the target layer through $B^{h.\ell}$. Also, in most cases, the performance of ML-BNMTF is comparable or superior to the performance of the third baseline method. So, it can be inferred that ML-BNMTF can adapt to variations of correlation of linkage between layers in these duplex networks and use it to improve the performance of link prediction.

It is also worth noting that ML-BNMTF is able to rank the communities of an auxiliary layer with respect to their relevance to the target layer. Ranking can be done using the diagonal values of $B^{h.\ell}$. This is more helpful when the local link overlap is not enough to explain the complex correlations between layers.

## VI. Conclusion

Our observations on real multiplex networks indicate that inter-layer community overlap exists in real multiplex networks, but the extent of overlap may vary across different domains. Specifically, social multiplex networks show a higher level of overlaps compared to other domains like biological and technological domains. Besides, further analysis showed the significance of these observations concerning the null models. Therefore, it can be inferred that membership in a community of one layer contains information about membership in communities of another layer (the intuition behind using $U^{\ell}$).

In addition, we showed that the value of edge overlap between layers varies across different communities of one specific layer. This observation indicates that although the presence of links is correlated in layers, the extent of this correlation is not the same across different communities (the intuition behind using $B^{h.\ell}$).

Putting these elements together, we introduced ML-BNMTF which uses both community membership information of nodes in auxiliary layer and the contribution of these communities in the rebuilding of the target layer. Experiments on real networks showed that ML-BNMTF is effective in transferring useful information for finding missing links in a target layer, especially when the use of direct link overlap is not effective.

Also, it is worth noting that although ML-BNMTF can handle weighted and directed networks using asymmetric and augmented $B$ matrices, here, we have confined the use of this model to unweighted and undirected multiplex networks. Relaxing this limitation and doing experiments on real networks can be considered as future work.